\documentclass[pra,letterpaper,twocolumn,showpacs]{revtex4-1} 
\usepackage{times,xspace} 
\usepackage{amsbsy,amssymb,amsmath,bm,bbold} 
\usepackage{graphicx,color,epsfig,rotate} 
\usepackage{fancyhdr} 
\usepackage{epstopdf}
\usepackage{graphicx}

\usepackage{float}
\usepackage[colorlinks=true,linkcolor=blue,citecolor=blue,urlcolor=blue]{hyperref}

\def\bbbc{{\mathchoice {\setbox0=\hbox{$\displaystyle\rm C$}\hbox{\hbox 
to0pt{\kern0.4\wd0\vrule height0.9\ht0\hss}\box0}} 
{\setbox0=\hbox{$\textstyle\rm C$}\hbox{\hbox 
to0pt{\kern0.4\wd0\vrule height0.9\ht0\hss}\box0}} 
{\setbox0=\hbox{$\scriptstyle\rm C$}\hbox{\hbox 
to0pt{\kern0.4\wd0\vrule height0.9\ht0\hss}\box0}} 
{\setbox0=\hbox{$\scriptscriptstyle\rm C$}\hbox{\hbox 
to0pt{\kern0.4\wd0\vrule height0.9\ht0\hss}\box0}}}}

\begin{document} 
	
	\title[Gandhi et.al.]{Higher order topological degeneracies and towards unique successive state-switching in a four-level open system}
	
	\author{Sayan Bhattacherjee, Harsh K. Gandhi, Arnab Laha,}

\author{Somnath Ghosh}
\email{somiit@rediffmail.com}
\affiliation{Department of Physics, Indian Institute of Technology Jodhpur, Rajasthan-342037, India}

\begin{abstract} 
	The physics of topological singularities, namely exceptional points (EPs), has been a key to wide range of intriguing and unique physical effects in non-Hermitian systems. In this context, the mutual interactions among four coupled states around fourth-order EPs (EP4s) are yet to be explored. Here we report a four-level parameter-dependent perturbed non-Hermitian Hamiltonian, mimicking quantum or wave-based systems, to explore the physical aspects of an EP4 analytically as well as numerically. The proposed Hamiltonian exhibit different orders of interaction schemes with the simultaneous presence of different higher-order EPs. Here an EP4 has been realized by mutual interaction between four coupled states with proper parameter manipulation. We comprehensively investigate the dynamics of corresponding coupled eigenvalues with stroboscopic parametric variation in the vicinity of the embedded EP4 to establish a new successive state-switching phenomenon among them; which proves to be robust even in the presence of different order of EPs. Implementing the relation of the perturbation parameters with the coupling control parameters, we exclusively report a region to host multiple EP4 in a specific system. The chiral behaviour of successive state-exchange has also been established near EP4. Proposed scheme enriched with physical aspects of EP4s should provide a new light manipulation tool in any anisotropic multi-state integrated system.  
\end{abstract} 
 
 
\maketitle %

\section{Introduction}

Beyond the conservative Hermitian quantum systems, the non-conservative or dissipative systems always present a richer physical impact as they exchange energies with the environment \cite{Moiseyev11}. Here, non-Hermitian formulations in quantum mechanics provide a better platform to understand the interaction between the energy-states of such open systems with their surrounding environments. During state-interactions in a parameter-dependent open system, the spectral degeneracies can be realized with the presence of branch-point singularities in the parameter space. An {\it Exceptional Point} (EP) of the order $N$ (say, EP{\it N}) is a special kind of topological singularity in system parameter space of non-Hermitian systems in general, for which $N$ number of eigenvalues and their corresponding eigenvectors simultaneously coalesce, and the effective Hamiltonian of the underlying system becomes defective \cite{Kato95,Heiss12}. Thus, a second-order EP (say, EP2) refers to a particular singularity where two coupled eigenvalues coalesce \cite{Sannino90,Heiss00}. In a very similar way, a third-order EP (say, EP3) can be realized with the coalescence of three coupled states \cite{Heiss08,Demange12,Heiss16}; however, there are several reports on EP3 where similar physical consequences can be achieved by winding around two EP2s associated with three interacting states \cite{Ryu12,Bhattacherjee19_1,Bhattacherjee19_2}.

In presence of EPs, the exotic physical phenomena have been widely investigated in a wide range of open systems like atomic \cite{Cartarius07,Menke16} and molecular \cite{Lefebvre09} spectrum, microwave cavities \cite{Dietz11}, Bose-Einstein \cite{Gutohrlein13} and Bose-Hurburrd \cite{Graefe08} systems, etc. Apart from these non-optical systems, the unconventional physical aspects of EPs have been mainly studied in various photonic systems like lasers \cite{Hodaei16,Wong16}, optical microcavities \cite{Laha17_1,Laha17_2,Laha19} and planar \cite{Ghosh16,Laha18} and coupled \cite{ZhangX18,ZhangX19,Schnabel17} waveguides, photonic crystals \cite{Ding15,Bykov18}, etc. Using optical gain-loss as nonconservative elements, such photonic systems provide a leading platform to meet a wide range of contemporary technological applications like unidirectional light transmission \cite{Yin13}, topological energy transfer \cite{Xu16}, asymmetric mode switching/conversion \cite{Ghosh16,Laha18,ZhangX18,ZhangX19}, resonance scattering \cite{Heiss15}, cross-polarization mode coupling \cite{Bykov18,Midya17}, lasing and antilasing \cite{Hodaei16,Wong16}, ultra-sensitive optical sensing \cite{Wiersig14,Wiersig16,Hodaei17,Chen17}, optical isolation with enhanced nonreciprocal effect \cite{Thomas16,Choi17}, stopping of light \cite{Goldzak18} etc. Recently, EPs have been also explored in cavity-optomechanics \cite{Jing17} in the context of phonon-magnon coupling \cite{Gao17} and phonon-lasing \cite{Lu17}. In various $\mathcal{PT}$-symmetric system, EPs have been studied in connection with broken $\mathcal{PT}$-symmetry \cite{Ganainy18}. For detail review, see ref. \cite{Miri19}.

The presence of an EP in parameter space unexpectedly modifies the dynamics of the system. A stroboscopic variation of control parameters enclosing an EP results in the permutation between the coupled states where they successively exchange their identities \cite{Heiss00,Bhattacherjee19_1,Bhattacherjee19_2,Cartarius07,Menke16,Laha17_1,Laha17_2,Laha19,Ghosh16,Laha18,Schnabel17}. This state-exchange phenomenon around a branch-point singularity is the fundamental proof of the exceptional behavior of that singularity in the sense that the singularity must behave like an EP. Such effect of parametric encirclement around an EP2 and corresponding topological properties \cite{Dembowski01} have been experimentally demonstrated for the first time in an microwave cavity \cite{Dembowski04}. During permutation between two coupled states, one of the corresponding eigenvectors acquires an additional Berry phase \cite{Lee SY12}. The successive state-flipping between three coupled states around an EP3 and their corresponding geometric phase behavior has been analytically established \cite{Bhattacherjee19_1,Bhattacherjee19_2,Lee SY12}, and also numerically demonstrated in a coupled waveguide system \cite{Schnabel17}. Instead of such stroboscopic parametric encirclement around EPs, for device-level implementation, if we consider time or analogous length scale dependent parametric variation to encircle an EP dynamically then adiabaticity of the system breaks down which essentially enables a nonadiabatic evolution of one of the two coupled states \cite{Gilary13}. In that case, only the eigenstate that evolves with lower average loss behaves adiabatically, and depending on the direction of rotation a specific eigenstate dominates at the end of encirclement process. Such a competition between the effect of EP and the adiabatic theorem leads to an asymmetric state-transfer phenomena \cite{Ghosh16,Laha18,ZhangX18,ZhangX19}.

 The cube root response near an EP3 hauls more complex physics in comparison with the square root response near an EP2. For example, if we consider EP-aided sensing application, then sensitivity can be immensely enhanced, exploiting an EP3 \cite{Hodaei17} in comparison with EP2 \cite{Wiersig14,Wiersig16}. So, it would be indeed quite interesting, if one can manipulate the mutual interaction between four coupled states simultaneously, then a fourth-order EP (say, EP4) can be encountered which could be suitable to study the even more complex physics of fourth-root response near the EP4. At an EP4, four coupled states should be analytically connected. However, with proper parameter manipulation, the simultaneous interaction among four coupled states around an EP4 has never been explored.

In this paper, we explore the analytical framework and corresponding topological properties of an EP4 for the first time. To study the state-dynamics alongside an EP4, we realize an open system, having four decaying eigenstates, that are subjected to a parameter dependent perturbation. We judiciously choose some control parameters to connect the passive system to the perturbation in such a way that we can simultaneously study different orders of interaction phenomena. With proper parameter manipulation, we encounter a situation where four coupled states are mutually interacting around a fourth-order singularity. Encircling this singularity in the system parameter plane, we explore an exclusive state-flipping phenomenon for the first time. Here four coupled states exchange their identities successively which confirms the presence of an EP4. In addition to EP4, we also explore the simultaneous existence of EP2s and EP3s in the same system and establish the possibility of the simultaneous existence of different orders of EPs in a particular system. Similar to 1$D$ exceptional-line connects which connects multiple number of EP2s \cite{Laha17_1,Laha17_2}, we corroborate the relation of the perturbation parameters with the coupling control parameters, we formulate a 3$D$ EP4-region within which multiple locations that could be labeled as EP4 coexist. The chiral behavior of state-exchange around the E4 has also been established. In this context, the findings are reported for the first time. Proposed scheme may be implemented using suitable state-of-the-art techniques in an anisotropic multi-state optical system.

\section{Mathematical modeling}

In order to achieve our goal, we consider a simple generic $4\times4$ non-Hermitian Hamiltonian matrix $\mathcal{H}$ having the form $H_0+\lambda H_p$.
\begin{equation}
\mathcal{H}=\left(\begin{array}{cccc}\widetilde{\varepsilon}_1 & 0 & 0 & 0 \\0 & \widetilde{\varepsilon}_2 & 0 & 0 \\0 & 0 & \widetilde{\varepsilon}_3 & 0\\0 & 0 & 0 & \widetilde{\varepsilon}_4\end {array}\right)+\lambda\left(\begin{array}{cccc}0 & \omega_p  & 0 & \omega_q \\ \omega_p & 0 & \omega_r & 0 \\ 0 & \omega_r  & 0 & \omega_s\\ \omega_q & 0 & \omega_s & 0\end {array}\right).
\label{equation_H}
\end{equation}

Here, the passive Hamiltonian $H_0$ is subjected to a parameter dependent complex perturbation $H_p$.  $\lambda$ represents a complex tunable parameter: $\lambda=\lambda_R+i\lambda_I$. $H_0$ consists of four complex states $\widetilde{\varepsilon}_j\,(j=1,2,3,4)$. Here, we consider $\widetilde{\varepsilon}_j=\varepsilon_j+i\tau_j\,(\tau_j<<\varepsilon_j)$ given that $\tau_j$ are the decay rates of the respective $\varepsilon_j$. The $H_p$ is parametrized by four interconnected perturbation parameters $\omega_p$, $\omega_q$, $\omega_r$ and $\omega_s$. Now, four eigenvalues of $\mathcal{H}$, say $E_j\,(j=1,2,3,4)$ are obtained by solving the eigenvalue-equation $|\mathcal{H}-EI|=0$ ($I\rightarrow\,4\times4$ identity matrix) which gives the quartic secular equation
\begin{equation}
E^4+p_1E^3+p_2E^2+p_3E+p_4=0;
\label{equation_E}
\end{equation}
where,
\begin{widetext}
\begin{subequations}
\begin{eqnarray}
 p_1=-\left(\widetilde{\varepsilon}_1+\widetilde{\varepsilon}_2+\widetilde{\varepsilon}_3+\widetilde{\varepsilon}_4\right)
\label{equation_p1}\\
 p_2=\widetilde{\varepsilon}_1\widetilde{\varepsilon}_2+\widetilde{\varepsilon}_2\widetilde{\varepsilon}_3+\widetilde{\varepsilon}_3\widetilde{\varepsilon}_4+\widetilde{\varepsilon}_4\widetilde{\varepsilon}_1+\widetilde{\varepsilon}_1\widetilde{\varepsilon}_3+\widetilde{\varepsilon}_2\widetilde{\varepsilon}_4-\lambda^2\left(\omega_p^2+\omega_q^2+\omega_r^2+\omega_s^2\right)\label{equation_p2}\\
p_3=-\left(\widetilde{\varepsilon}_1\widetilde{\varepsilon}_2\widetilde{\varepsilon}_3+\widetilde{\varepsilon}_2\widetilde{\varepsilon}_3\widetilde{\varepsilon}_4+\widetilde{\varepsilon}_1\widetilde{\varepsilon}_3\widetilde{\varepsilon}_4+\widetilde{\varepsilon}_1\widetilde{\varepsilon}_2\widetilde{\varepsilon}_4\right)
	+\lambda^2\left\{(\widetilde{\varepsilon}_1+\widetilde{\varepsilon}_2)\omega_s^2+(\widetilde{\varepsilon}_2+\widetilde{\varepsilon}_3)\omega_q^2+(\widetilde{\varepsilon}_3+\widetilde{\varepsilon}_4)\omega_p^2+(\widetilde{\varepsilon}_4+\widetilde{\varepsilon}_1)\omega_r^2\right\}
\label{equation_p3}\\ p_4=\widetilde{\varepsilon}_1\widetilde{\varepsilon}_2\widetilde{\varepsilon}_3\widetilde{\varepsilon}_4-\lambda^2\left(\widetilde{\varepsilon}_1\widetilde{\varepsilon}_2\omega_s^2+\widetilde{\varepsilon}_2\widetilde{\varepsilon}_3\omega_q^2+\widetilde{\varepsilon}_3\widetilde{\varepsilon}_4\omega_p^2+\widetilde{\varepsilon}_4\widetilde{\varepsilon}_1\omega_r^2\right)-\lambda^4\left(\omega_p\omega_s+\omega_q\omega_r\right)^2.
\label{equation_p4}
\end{eqnarray}	
\end{subequations}
Using the Ferrari's method~\cite{Boyer91}, the roots of the Eq.~\ref{equation_E} can be written as
\begin{subequations}\begin{eqnarray}
E_{1,2}=-\frac{p_1}{4}-\eta\pm\frac{1}{2}\sqrt{-4\eta^2-2m_1+\frac{m_2}{\eta}},\label{equation_E12}
\\
E_{3,4}=-\frac{p_1}{4}+\eta\pm\frac{1}{2}\sqrt{-4\eta^2-2m_1-\frac{m_2}{\eta}};\label{equation_E34}
\end{eqnarray}
\end{subequations}
where
\begin{equation}
\eta=\frac{1}{2}\sqrt{-\frac{2}{3}m_1+\frac{1}{3}\left(\kappa+\frac{m_3}{\kappa}\right)}\quad\textnormal{with}\quad\kappa=\left(\frac{m_4+\sqrt{m_4^2-4m_3^3}}{2}\right)^{1/3}. 
\end{equation}
\end{widetext}
Here,
\begin{subequations}
\begin{eqnarray}
m_1=-\frac{3p_1^2}{8}+p_2,
\label{equation_m1}\\
m_2=\frac{p_1^3}{8}-\frac{p_1p_2}{2}+p_4,
\label{equation_m2}\\
m_3=p_2^2-3(p_1p_3+4p_4),
\label{equation_m3}\\
m_4=2p_2^3-9p_2(p_1p_3+8p_4)+27(p_1^2p_4+p_3^2).
\label{equation_m4}
\end{eqnarray}
\end{subequations}

Thus, the roots of Eq. \ref{equation_E} given by Eqs. \ref{equation_E12} and \ref{equation_E34} represent the eigenvalues of $\mathcal{H}$. Now to control the couplings between $E_j$, we introduce a new parameter $\delta$ to modulate the interconnected perturbation parameters. Here, we customize the perturbation parameters in terms of $\delta$ as
\begin{subequations}
\begin{eqnarray}
\omega_p=4\delta-10^{-4},\,\,\omega_q=\delta-0.1,\,\, \label{equation_omega1} \\
\omega_r=0.95-\delta/2,\,\,\textnormal{and}\,\,\omega_s=0.5-\delta.
\label{equation_omega2}
\end{eqnarray}
\end{subequations}
Thus, with the simultaneous variation of complex $\lambda\,(=\lambda_R+i\lambda_I)$ and $\delta$, the perturbation parameters $\omega_k\,(k=p,q,r,s)$ control the interactions between $E_j\,(j=1,2,3,4)$. Using this framework, various interactions phenomena are described in the following section. During optimization, we choose the passive eigenvalues $\varepsilon_1=0.9$, $\varepsilon_2=0.8$, $\varepsilon_3=1.25$ and $\varepsilon_4=0.25$ with the corresponding decay rates $\tau_1=5\times10^{-3}$, $\tau_2=2.5\times10^{-3}$, $\tau_3=0.2\times10^{-3}$ and $\tau_4=0.01\times10^{-3}$. Here we consider $\tau_j<<\varepsilon_j$ to implement this analytical model on any feasible anisotropic prototype device.

\section{Different order of interactions between the coupled states}
\label{sec3}

With consideration of the specific optimized values as described in the previous section, we study the interactions between $E_j\,(j=1,2,3,4)$,to which we simultaneously vary the complex $\lambda$ and $\delta$ within judiciously chosen regions. Here $\lambda_R$ varies within the range from $5.6$ to $5.7$, whereas $\lambda_I$ varies simultaneously maintaining the ratio $\lambda_I/\lambda_R=-10^{-3}$. The choice of $\lambda_I$ of the order $10^{-3}$ of $\lambda_R$ makes the physical system to be more realistic. The span for the variation of $\delta$ has been chosen in between $[-0.04, 0.04]$.
\begin{figure*}[htbp]
	\centering
	\includegraphics[width=14cm]{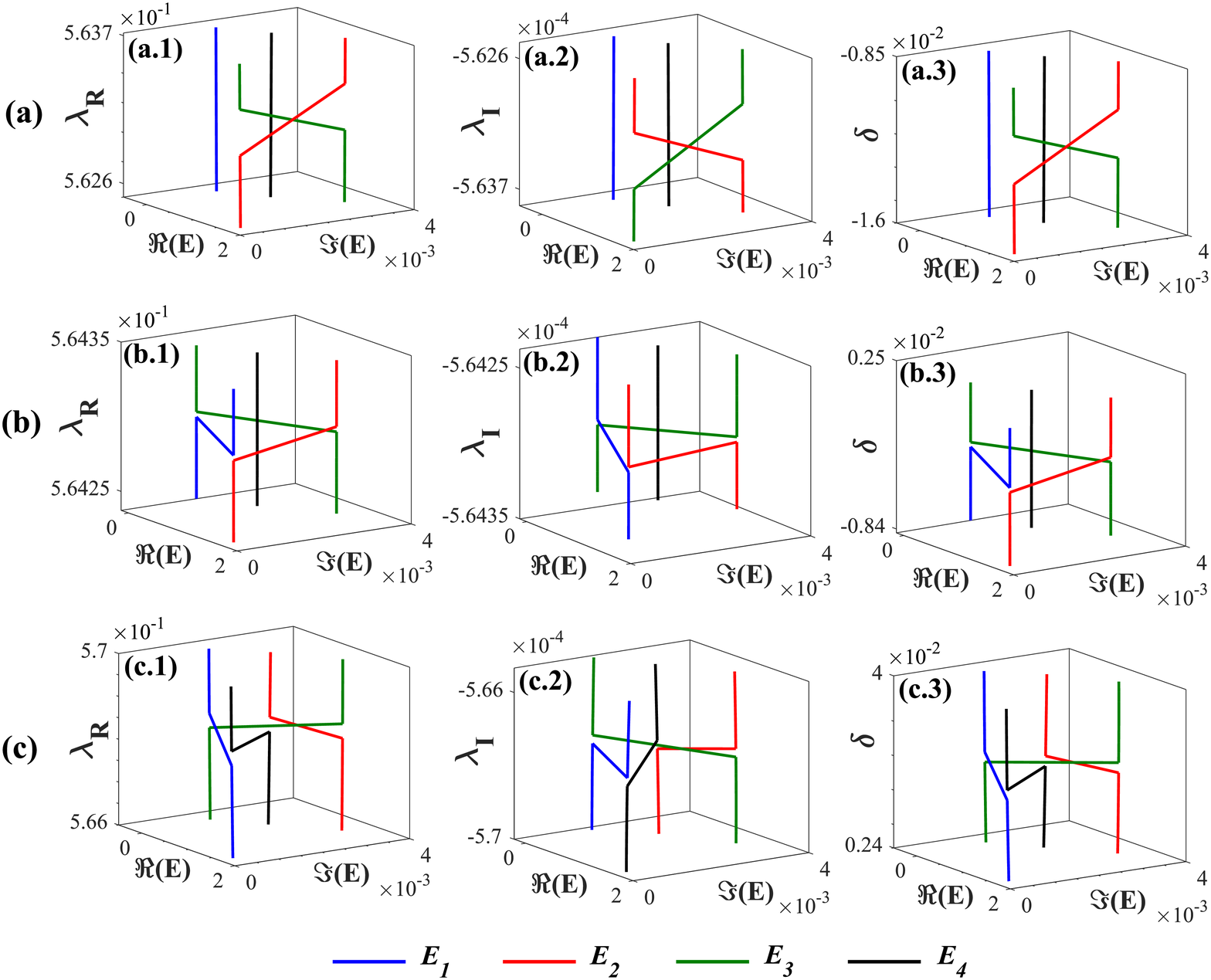}
	\caption{{\bf Interactions between the eigenvalues with the variations of the perturbation parameters}. Trajectories of $E_{j}\,(j=1,2,3,4)$ have been shown by blue, red, green and black lines, respectively. \textbf{(a)} A second-order interaction between $E_1$ and $E_2$, unaffecting $E_3$ and $E_4$, with respect to (a.1) $\lambda_R$, (a.2) $\lambda_I$ and (a.3) $\delta$. \textbf{(b)} A third-order interaction between $E_1$, $E_2$ and $E_3$, unaffecting $E_4$, with respect to (b.1) $\lambda_R$, (b.2) $\lambda_I$ and (b.3) $\delta$. \textbf{(c)} A fourth-order interaction between $E_1$, $E_2$, $E_3$ and $E_4$ with respect to (c.1) $\lambda_R$, (c.2) $\lambda_I$ and (c.3) $\delta$}       
	\label{fig1}
\end{figure*}

We study the interactions between $E_j\,(j=1,2,3,4)$ in Fig. \ref{fig1} with an increasing $\lambda$ and $\delta$ within the chosen limits. The trajectories of $E_{j}\,(j=1,2,3,4)$ have been shown using blue, red, green and black lines, respectively. In Fig. \ref{fig1}(a), we can observe the interaction between $E_1$ and $E_2$ in a certain range of the control parameters (within the specified span), where $E_3$ and $E_4$ remain unaffected. The equivalent dynamics of $E_j\,(j=1,2,3,4)$ with respect to $\lambda_R$, $\lambda_I$ and $\delta$ have been depicted in Figs. \ref{fig1}(a.1), (a.2) and (a.3), respectively. Thus from the interaction phenomenon shown in Fig. \ref{fig1}, it can be inferred that there should be a singularity of second-order near the interaction regime of $E_1$ and $E_2$. Now for further increase in $\lambda$ and $\delta$, we observe the simultaneous interaction between $E_1$, $E_2$ and $E_3$, unaffecting $E_4$, which is shown in Fig. \ref{fig1}(b). Here the similar behavior of $E_{j}\,(j=1,2,3,4)$ concerning $\lambda_R$, $\lambda_I$ and $\delta$, as can be seen in Figs.\ref{fig1}(b.1), (b.2) and (b.3), respectively, endorse the presence of a third order singularity in ($\lambda,\delta$)-plane. Now, after investigating the second and third oder interactions, we further increase the values of the control parameters to study the fourth-order interaction; which has been shown in Fig. \ref{fig1}(c). Here we observe that for comparably higher values of $\lambda$ and $\delta$, all the four states $E_{j}\,(j=1,2,3,4)$ are mutually interacting and show the similar coupling natures with respect to $\lambda_R$, $\lambda_I$ and $\delta$, as depicted in Figs.\ref{fig1}(c.1), (c.2) and (c.3). Such mutual coupling between four interacting states confirms about the presence of a singularity of the fourth order in the system parameter plane.

\begin{figure*}[htpb]
	\centering
	\includegraphics[width=15cm]{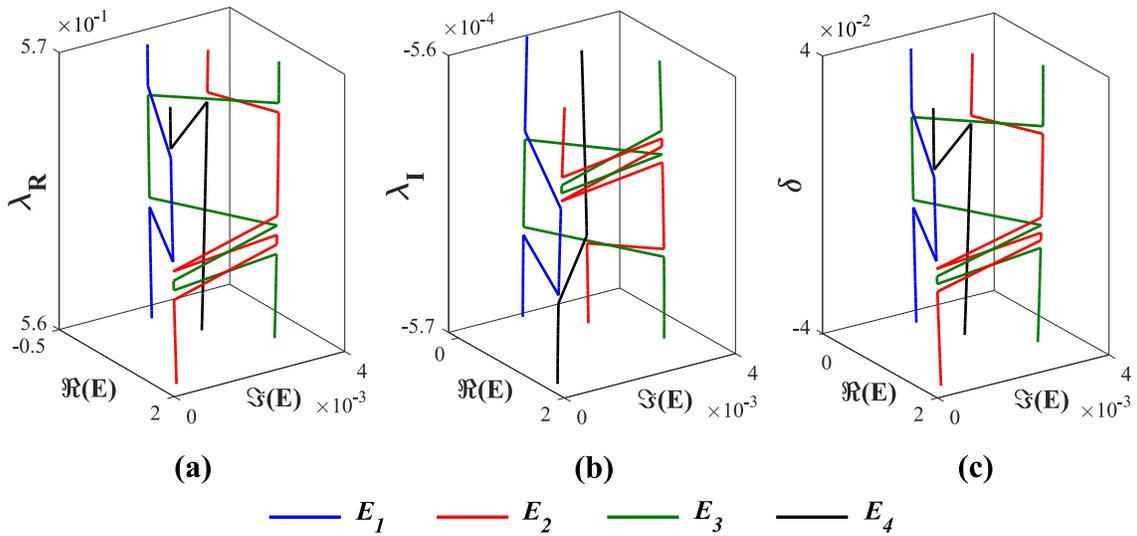}
	\caption{Simultaneous representation of different orders of interactions shown in Fig. \ref{fig1} with respect to \textbf{(a)} $\lambda_R$, \textbf{(b)} $\lambda_I$ and \textbf{(c)} $\delta$.}
	\label{fig2}
\end{figure*}

All kind of interaction phenomena of different orders (as shown in Fig. \ref{fig1}) which are hosted by the Hamiltonian $\mathcal{H}$ (given by Eq. \ref{equation_H}) have been simultaneously presented in Fig. \ref{fig2} for the entire chosen span of $\lambda$ and $\delta$. In Figs. \ref{fig2}(a), (b) and (c), the overall interaction phenomena among $E_j$ have been shown with respect to $\lambda_R$, $\lambda_I$ and $\delta$, respectively; where we observe that at the initial point of the chosen scale of the control parameters, the eigenvalues remain noninteracting, and then with increase in parametric values, they exhibit different order of interactions for different parametric regions. Thus identifying these particular regions in the parameter plane, we can realize the presence of singularities of different orders. In the following section, we examine the exceptional behavior of the embedded singularities by moving around them in system parameter plane.   

\section{Physical effects of topological singularities: Toward successive state-switching}

If a singularity behaves like an EP, then its presence inside a closed parameter space of the underlying system leads to significant modifications in the dynamics of the corresponding coupled states due to the influence of the coupling parameters. Quasi-statically encircling an EP in parameter space results in the permutation between the coupled eigenvalues. Around an EP, the corresponding coupled eigenvalues exchange their identities adiabatically. To enclose the singularities, we use the following parametric equation in ($\lambda,\delta$)-plane.
\begin{subequations}
\begin{eqnarray}
\lambda_R(\phi)=a_0\left[1+r_1\,\cos(\phi)\right] \,\, \\ \,\, \delta(\phi)=b_0\left[1+r_2\,\sin(\phi)\right].
\label{equation_circle} 
\end{eqnarray}
\end{subequations}

Here ($a_0,b_0$) represent the center of the parametric loop. $r_1$ and $r_2$ are two characteristics parameters to control the variations of $\lambda_R$ and $\delta$ over a tunable angle $\phi$ given that $\phi\in[0,2\pi]$. To encircle a singularity, we choose the variation of $\lambda_R$ and $\delta$ in Eq. \ref{equation_circle}, where there is a variation of $\lambda_I$ maintaining the ratio $\lambda_I/\lambda_R=-10^{-3}$ (as mentioned in section \ref{sec3}). Such overall parameter space ($\lambda_R,\,\lambda_I\,\textnormal{and}\,\delta$) variation significantly affect the dynamics of the coupled states. Judiciously choosing the characteristics parameters of Eq. \ref{equation_circle}, we can encircle the single or multiple singularities (even the singularities having the different orders) to scan the enclosed area. Now if we establish the successive state switching by parametric encirclement around the embedded singularities of different orders, then we can confirm that the proposed Hamiltonian $\mathcal{H}$ hosts different order of EPs \cite{Heiss00,Bhattacherjee19_1,Bhattacherjee19_2,Cartarius07,Menke16,Laha17_1,Laha17_2,Laha19,Ghosh16,Laha18,Schnabel17}.

\begin{figure*}[htpb]
	\centering
	\includegraphics[width=15cm]{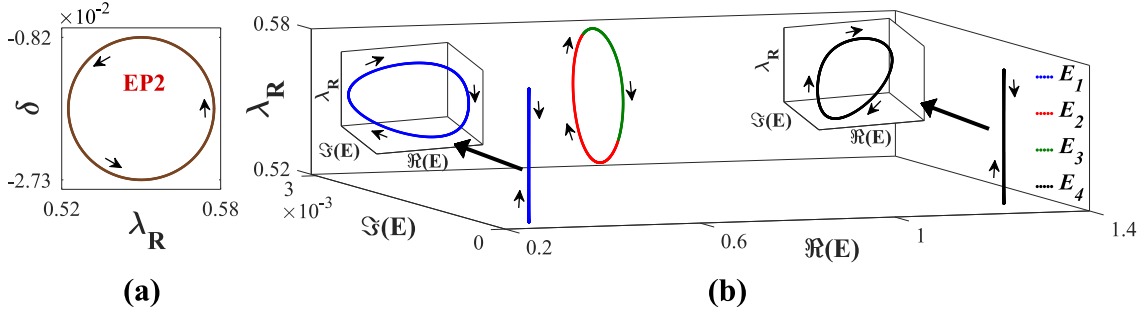}
	\caption{{\bf State-flipping between a pair of coupled states}. \textbf{(a)} Encircling an EP2 in ($\lambda_R$,$\delta$)-plane. \textbf{(b)} Corresponding dynamics of $E_j\,(j=1,2,3,4)$ in complex $E$-plane with respect to $\lambda_{R}$ showing the flipping between the coupled $E_1$ and $E_2$. Trajectories of $E_3$ and $E_4$ have been zoomed in the respective insets for proper visualization. Arrows in both (a) and (b) indicate the direction of progression.}
	\label{fig3}
\end{figure*}

Now, we predict the approximate second-order interaction region between $E_1$ and $E_2$ (except $E_3$ and $E_4$), as shown in Fig. \ref{fig1}(a), and judiciously choose $a_0=0.55,\,b_0=-0.0178,\,r_1=0.05,\,\textnormal{and}\,r_2=0.45$ to enclose the associated second order singularity. The corresponding parametric loop has been shown in Fig. \ref{fig3}(a). Looking at the ranges of $x$-axis and $y$-axis of Fig. \ref{fig3}(a) and $y$-axes of Figs. \ref{fig1}(a.1) and (a.3), we can confirm that the described parametric loop in Fig. \ref{fig3}(a) perfectly encloses the associated singularity that is responsible for the coupling between $E_1$ and $E_2$. Following a very slow evolution along this parametric loop, we plot the corresponding trajectories of $E_j\,(j=1,2,3,4)$ in Fig. \ref{fig3}(b). Here we have shown that for one complete cycle around the singularity in parameter plane, the coupled eigenvalues $E_1$ and $E_2$ are permuted by exchanging their identities and make a complete loop in complex eigenvalue plane, whereas the unaffected states $E_3$ and $E_4$ remain itself to make individual loops. The trajectories of $E_3$ and $E_4$ have been zoomed in the corresponding insets in Fig. \ref{fig3}(b). Such unconventional state-dynamics in the complex eigenvalue plane proves that the identified second-order singularity between $E_1$ and $E_2$ behaves as an EP2 \cite{Heiss00,Cartarius07,Menke16,Laha17_1,Laha17_2,Laha19,Ghosh16,Laha18}. In Fig. \ref{fig3}(b), we have shown the state-dynamics in the complex-eigenvalue plane concerning the parameter $\lambda_R$, however, similar state dynamics can also be observed with respect to the parameters $\lambda_I$ and $\delta$.  

\begin{figure*}[htpb]
	\centering
	\includegraphics[width=15cm]{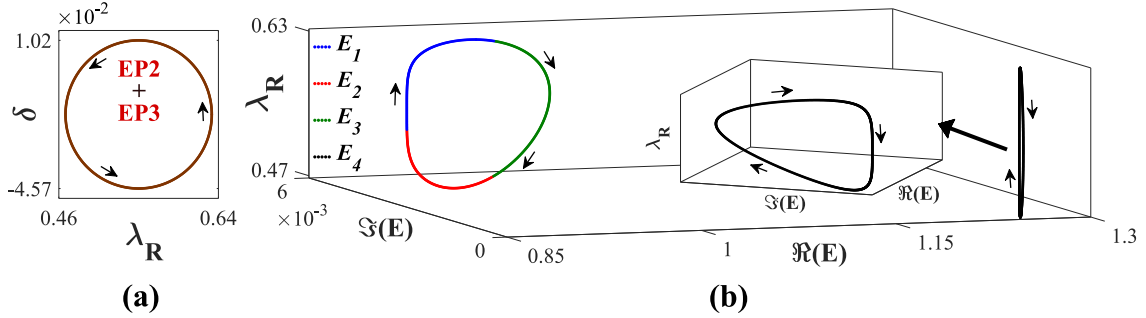}
	\caption{{\bf Successive state-flipping between three coupled states}. \textbf{(a)} Parametric variation enclosing an EP2 and an EP3 in ($\lambda_R$,$\delta$)-plane. \textbf{(b)} Corresponding dynamics of $E_j\,(j=1,2,3,4)$ in complex $E$-plane with respect to $\lambda_{R}$ showing the successive flipping between the coupled $E_1$, $E_2$ and $E_3$. Trajectory of $E_4$ has been zoomed in the inset for proper visualization. Arrows in both (a) and (b) indicate the direction of progression.}
	\label{fig4}
\end{figure*}

After exploring the EP2 in the proposed system, we look into the parametric region where $E_1$, $E_2$ and $E_3$ (except $E_4$) are mutually coupled (as shown in Fig. \ref{fig1}(b)). To enclose this region, we perform an encirclement process by choosing the characteristics parameters of Eq. \ref{equation_circle} as $a_0=0.55,\,b_0=-0.028,\,r_1=0.15,\,\textnormal{and}\,r_2=0.55$. These parameters are chosen in such a way that the resulting parameter space, shown in Fig. \ref{fig4}(a), encloses the third-order singularity in addition to the EP2 (between $E_1$ and $E_2$ only; as described in Fig. \ref{fig3}). Thus, we can examine the effects of the third-order singularity even in presence of a different lower order singularity. Now following an quasi-static encirclement process along the closed loop shown in Fig. \ref{fig4}(a), we plot the dynamics of $E_j\,(j=1,2,3,4)$ in Fig. \ref{fig4}(b) concerning the parameter $\lambda_R$ (however, instead of $\lambda_R$, we can also choose $\lambda_I$ or $\delta$). Here, three coupled eigenvalues $E_1$, $E_2$ and $E_3$ flip successively by exchanging their identities adiabatically in complex eigenvalue plane for one complete loop in parametric plane. However, the non-interacting state $E_4$ is not affected by the dynamics of other three states and keeps its self-identity by making an individual loop. The magnified view of the trajectory of $E_4$ has been shown in the inset. Such state dynamics in the complex eigenvalue plane, as shown in \ref{fig4}(b), following the parameter space, as shown in \ref{fig4}(a), clearly justify that the enclosed third-order singularity between $E_1$, $E_2$ and $E_3$ behaves as an EP3 \cite{Bhattacherjee19_1,Bhattacherjee19_2,Schnabel17}. Here the exotic effect of the identified EP3 on the state-dynamics is robust even in presence of an EP2 inside the parametric loop. If we choose a similar parameter space that only encircle the approximate position of EP3, even then one should observe the similar state-dynamics in the complex eigenvalue-plane.       

\begin{figure*}[htpb]
	\centering
	\includegraphics[width=15cm]{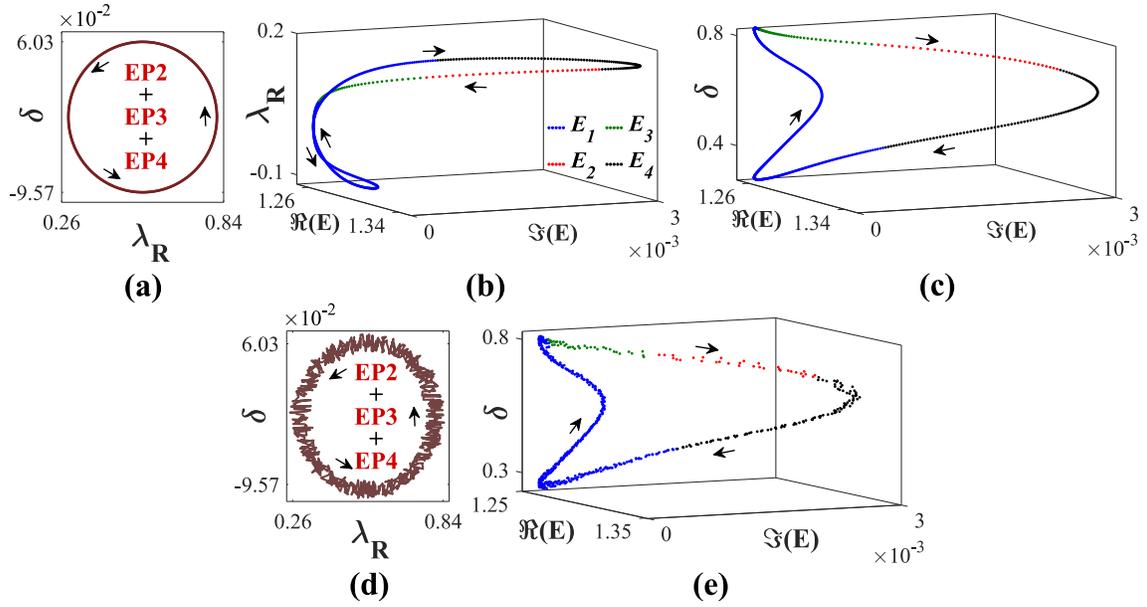}
	\caption{{\bf Successive state-flipping between four coupled states}. \textbf{(a)} Parametric encirclement around an EP2, an EP3 and an EP4 in ($\lambda_R$,$\delta$)-plane. Corresponding dynamics of $E_j\,(j=1,2,3,4)$ in complex $E$-plane with respect to \textbf{(b)} $\lambda_{R}$ and \textbf{(c)} $\delta$ showing the successive flipping between all the coupled states. \textbf{(d)} Similar parametric encirclement as shown in (a) with additional fluctuation. \textbf{(e)} Corresponding dynamics of $E_j\,(j=1,2,3,4)$ in complex $E$-plane with respect to $\delta$. Arrows in (a)--(e) indicate the direction of progressions.}
	\label{fig5}
\end{figure*}

Successfully verifying the topological properties of an EP2 and an EP3, then, we study the dynamics of the proposed four-level Hamiltonian $\mathcal{H}$ (Eq. \ref{equation_H}). We encircle the approximate position of the embedded fourth-order singularity where all the supported four states $E_j\,(j=1,2,3,4)$ are analytically connected. Accordingly, we choose the characteristics parameters of Eq. \ref{equation_circle} as $a_0=0.55,\,b_0=-0.019,\,r_1=0.5,\,\textnormal{and}\,r_2=3.4$. Such set of parameters also give the opportunity to study the immutable behaviour of the fourth-order singularity even in presence of encountered EP2 and EP3. The chosen parametric contour has been shown in Fig. \ref{fig5}(a). In Fig. \ref{fig5}(b), we study the corresponding dynamics of $E_j\,(j=1,2,3,4)$ following a quasi-static parametric variation along the loop described in Fig. \ref{fig5}(a). As shown in Fig. \ref{fig5}(b), following one complete parametric cycle, all the coupled eigenvalues successively exchange their identities and make a complete loop in the complex eigenvalue plane. Here the state-dynamics have been shown with respect to the parameter $\lambda_R$. In Fig. \ref{fig5}(c), we have shown the similar successive state-flipping phenomena concerning the parameter $\delta$, for the exactly same parametric loop. Such state dynamics confirms the exceptional nature of the embedded fourth-order singularity as an EP4, i.e. fourth-order EP. Thus we have successfully explored an exclusive state-exchange among the four coupled states around an EP4 for the first time. We can also observe that there is no effect of EP2 and EP3 on the state dynamics, if an EP4 is properly enclosed in the system parameter space. During implementation of the proposed scheme in any realistic system, some unwanted tolerance may be appeared during the parametric encirclement process. To take into account such fabrication tolerances, we add some random fluctuations (up to $\sim10\%$) on variation of the parameters following the same parametric loop as shown in Fig. \ref{fig5}(a). The modified parametric loop and corresponding state-dynamics have been shown in Fig. \ref{fig5}(d) and (e), respectively. Investigating the state-dynamics as can be seen in Fig. \ref{fig5}(e), we can conclude that the successive state-flipping phenomena around an EP4 is robust even in presence of the parametric fluctuation.However, this is robust till the amount of fluctuation does not affect the approximate location of the EP4. 

Note that, to facilitate matter, we do not consider the explicit time dependence on the parametric variation for the proposed Hamiltonian $\mathcal{H}$ (given by Eq. \ref{equation_H}). Thus within this framework, the dynamical EP encirclement process and corresponding nonadiabatic chiral state-transfer phenomena can not be realized. In this work, we straightforwardly investigate the dynamics of four complex eigenvalues in the vicinity of an EP4 and investigate corresponding topological properties.

\section{Analytic picture of an EP4}
Here, we describe the analytic structure of the eigenvalues and the corresponding eigenfunctions near an EP4\cite{Heiss08}. To describe the peculiar nature of the four-fold coalescence in the Hamiltonian $\mathcal{H}\,(=H_0+\lambda H_p)$ (given by Eq. \ref{equation_H}), we consider a particular point $\lambda_c$, where four levels are analytically connected, and a {\it critical eigenvalue} $E_c$ at the coalescing point. Now to consider such four-level coalescence, the set of equations
\begin{equation}
\frac{\mathrm{d}^k}{\mathrm{d}E^{k}}\det\left|\mathcal{H}(\lambda)-EI\right|=0\qquad k=0,1,2,3
\label{a1}
\end{equation}
must be satisfied. Accordingly, we choose a general set of eigenvalues in terms of $\lambda_c$ and $E_c$ as
\begin{equation}
E_j(\lambda)=E_c+\sum_{l=1}^{\infty}a_l\left(\sqrt[4]{\lambda-\lambda_{c}}\right)^{l} \,\textnormal{with}\,\,j=1,2,3,4.
\label{a2}               
\end{equation}
Here, $a_l$ represent some real constants. $j=1,2,3,4$ represent the levels that are defined by the quantity $\left(\sqrt[4]{\lambda-\lambda_{c}}\right)$ on first, second, third and fourth Riemann sheet in the $\lambda$-plane. After expanding, Eq. \ref{a2} can be written more explicitly as
\begin{widetext}
\begin{equation}
E_{j}(\lambda)=E_{c}+\sum_{l=1}^{\infty}a_{l}\left[\sqrt[4]{|\lambda-\lambda_{c}|}\exp\left(\frac{i \arg\left(\lambda-\lambda_c\right)+2i\pi(j-1)}{4}\right)\right]^l
\label{a3}
\end{equation}
\end{widetext}
Now, the structure of the corresponding eigenfunctions can be written as
\begin{equation}
|\psi_{j}(\lambda)\rangle =|\psi_{\mathrm{EP4}}\rangle+\sum_{l=1}^{\infty}\left(\sqrt[4]{\lambda-\lambda_{c}}\right)^{l}|\phi_{k}\rangle
\label{a4}
\end{equation}
by considering a {\it critical eigenfunction} at EP4 as $|\psi_{\mathrm{EP4}}\rangle$. Explicitly we can further write the eigenfunctions corresponding to the Riemann sheet of the fourth-root as
\begin{equation}
|\psi_{j}(\lambda)\rangle =|\psi_{\mathrm{EP4}}\rangle+\sum_{l=1}^{\infty}\left(\sqrt[4]{\lambda-\lambda_{c}}\right)^{l}\left|\phi^j_{k}\right\rangle
\label{a5}
\end{equation}
with $\left|\phi^j_{k}\right\rangle=\exp\left(i \arg\left(\lambda-\lambda_c\right)/4+2i\pi(j-1)/4\right)\left|\phi_{k}\right\rangle$. Now, the all possible pairs of eigenfunctions given by Eq. \ref{a5} should be bi-orthogonal for all $\lambda\ne\lambda_c$ as

\begin{equation}
\left\langle\psi_i(\lambda)\big|\psi_j(\lambda)\right\rangle=N_j(\lambda)\delta_{i,j},\quad
\end{equation}
provided that
\begin{equation}
\quad \sum_{l}\frac{\left|\psi_{j}(\lambda)\right\rangle\left\langle\psi_{j}(\lambda)\right|}{\left\langle \psi_{j}(\lambda)\big|\psi_{j}(\lambda)\right\rangle}=I.
\label{a6}
\end{equation}

Now, if we replace one of the eigenfunctions of the product (given by Eq. \ref{a6}) with the critical eigenfunction $|\psi_{\mathrm{EP4}}\rangle$ then we can write 
\begin{equation}
\left\langle\psi_i(\lambda)\big|\psi_{\mathrm{EP4}}\right\rangle\sim\vartheta\left(\sqrt[4]{\lambda-\lambda_{c}}\right)^3\quad\textnormal{for}\quad\lambda\rightarrow\lambda_c
\label{a7}
\end{equation}
with $\vartheta$ as some constant. Thus, once we consider $\lambda\rightarrow\lambda_c$, $\left\langle\psi_{\mathrm{EP4}}\big|\psi_{\mathrm{EP4}}\right\rangle$ vanishes even $i\ne j$; which means the coalescence of the eigenvectors. In addition, if $|\phi_1\rangle$ is the associated with first power of $\left(\sqrt[4]{\lambda-\lambda_{c}}\right)$ then $\left\langle\psi_{\mathrm{EP4}}\big|\phi_1\right\rangle$ should also vanish.

Now around the EP4, we can write the $|\psi_{\mathrm{EP4}}\rangle$ as the linear combination of the coupled eigenvectors $|\chi_{j}(\lambda)\rangle$ with some constants $c_j$ like
\begin{subequations}
	\begin{eqnarray}
	|\psi_{\mathrm{EP4}}\rangle=\sum_{j=1}^{4}c_{j}(\lambda)|\chi_{j}(\lambda)\rangle\,\,\textnormal{with},\,\,\\
	|\chi_{j}(\lambda)\rangle=\frac{|\psi_{j}(\lambda)\rangle}{\sqrt{\left\langle\psi_i(\lambda)\big|\psi_j(\lambda)\right\rangle}}
	\label{a8}
	\end{eqnarray}
\end{subequations}

Here, the solutions of Eq. \ref{a8} while $\lambda\rightarrow\lambda_{c}$ would yield the basic structure of general eigenfunction with corresponding phase relations based upon the fourth roots of unity. The possible combinations of $c_j$ have been given below.
\begin{widetext}
\begin{eqnarray}
\left(\begin{array}{c}c_1(\lambda)\\c_2(\lambda)\\c_3(\lambda)\\c_4(\lambda)\end{array}\right)\sim\kappa_1\sqrt[4]{|\lambda-\lambda_{c}|}\left(\begin{array}{c}1\\e^{+i\pi/2}\\e^{i\pi}\\e^{-i\pi/2}\end{array}\right),\,\,\,\,\,\,\,\,\,\left(\begin{array}{c}c_1(\lambda)\\c_2(\lambda)\\c_3(\lambda)\\c_4(\lambda)\end{array}\right)\sim\kappa_2\sqrt[4]{|\lambda-\lambda_{c}|}\left(\begin{array}{c}e^{-i\pi/2}\\e^{+i\pi/2}\\1\\e^{i\pi}\end{array}\right)\nonumber\\
\left(\begin{array}{c}c_1(\lambda)\\c_2(\lambda)\\c_3(\lambda)\\c_4(\lambda)\end{array}\right)\sim\kappa_3\sqrt[4]{|\lambda-\lambda_{c}|}\left(\begin{array}{c}e^{+i\pi/2}\\e^{i\pi}\\1\\e^{-i\pi/2}\end{array}\right)\,\textnormal{and}\,\left(\begin{array}{c}c_1(\lambda)\\c_2(\lambda)\\c_3(\lambda)\\c_4(\lambda)\end{array}\right)\sim\kappa_4\sqrt[4]{|\lambda-\lambda_{c}|}\left(\begin{array}{c}e^{+i\pi/2}\\e^{-i\pi/2}\\e^{i\pi}\\1\end{array}\right)
\end{eqnarray}
\end{widetext}

We also obtain the fact that the diverging of $|\chi_{j}(\lambda)\rangle$ will lead to a finite valued $|\psi_{\mathrm{EP4}}\rangle$. Thus from Eq. \ref{a8}, we further conclude that 
\begin{equation}
\sum_{j=1}^{4}c_{j}(\lambda)=0;
\label{a9}
\end{equation} 
since, $\left\langle\overline{\psi}_{\mathrm{EP4}}\big|\psi_{\mathrm{EP4}}\right\rangle =0$. Eq. \ref{a9}, that resembles to the chirality condition around EP2, also leads the chiral nature of modes around an EP4.

\section{Formulation of a region to host multiple EP4: Exceptional Region}
In this section, we study the specific relations between the perturbation parameters (which are connected by a specific parameter $\delta$) and the independent coupling control parameter $\lambda$ to formulate a specific parametric region in which the fourth-order coupling can occur multiple times. This specific parametric region has been named as ``EP4-region'', i.e., this region can host multiple number of EP4s.

To describe such a region we have made some special settings in our proposed Hamiltonian $\mathcal{H}$ given in Eq. \ref{equation_H}. Initially, to facilitate the situation, we consider $\omega_s=1$ and rewrite the Eq. \ref{equation_H} as
\begin{equation}
\mathcal{H}\Bigg|_{\omega_s=1}=\left(\begin{array}{cccc}\widetilde{\varepsilon}_1 & \omega_p\lambda & 0 & \omega_q\lambda \\\omega_p\lambda & \widetilde{\varepsilon}_2 & \omega_r\lambda & 0 \\0 & \omega_r\lambda & \widetilde{\varepsilon}_3 & 0\\\omega_q\lambda & 0 & 0 & \widetilde{\varepsilon}_4\end {array}\right)+ 
 \lambda\left(\begin{array}{cccc}0 & 0 & 0 & 0 \\0 & 0 & 0 & 0 \\ 0 & 0  & 0 & 1\\ 0 & 0 & 1 & 0\end {array}\right)
\label{equation_H1}
\end{equation}

Such special consideration has been made to explore the relation of $\omega_p$ with $\omega_q$ and $\omega_r$ over the independent variation of $\lambda$. In this case Eq. \ref{equation_E} can be rewritten as
\begin{equation}
E^4+p_1E^3+p'_2E^2+p'_3E+p'_4=0;
\label{equation_Ea}
\end{equation}
where $p_1$ is given by Eq. \ref{equation_p1}, and $p'_2,\,p'_3\,\textnormal{and}\,p'_4$ are coming from Eqs. \ref{equation_p2}, \ref{equation_p3} and \ref{equation_p4}, respectively, considering the special setting $\omega_s=1$.
\begin{figure*}[htpb]
	\centering
	\includegraphics[width=15cm]{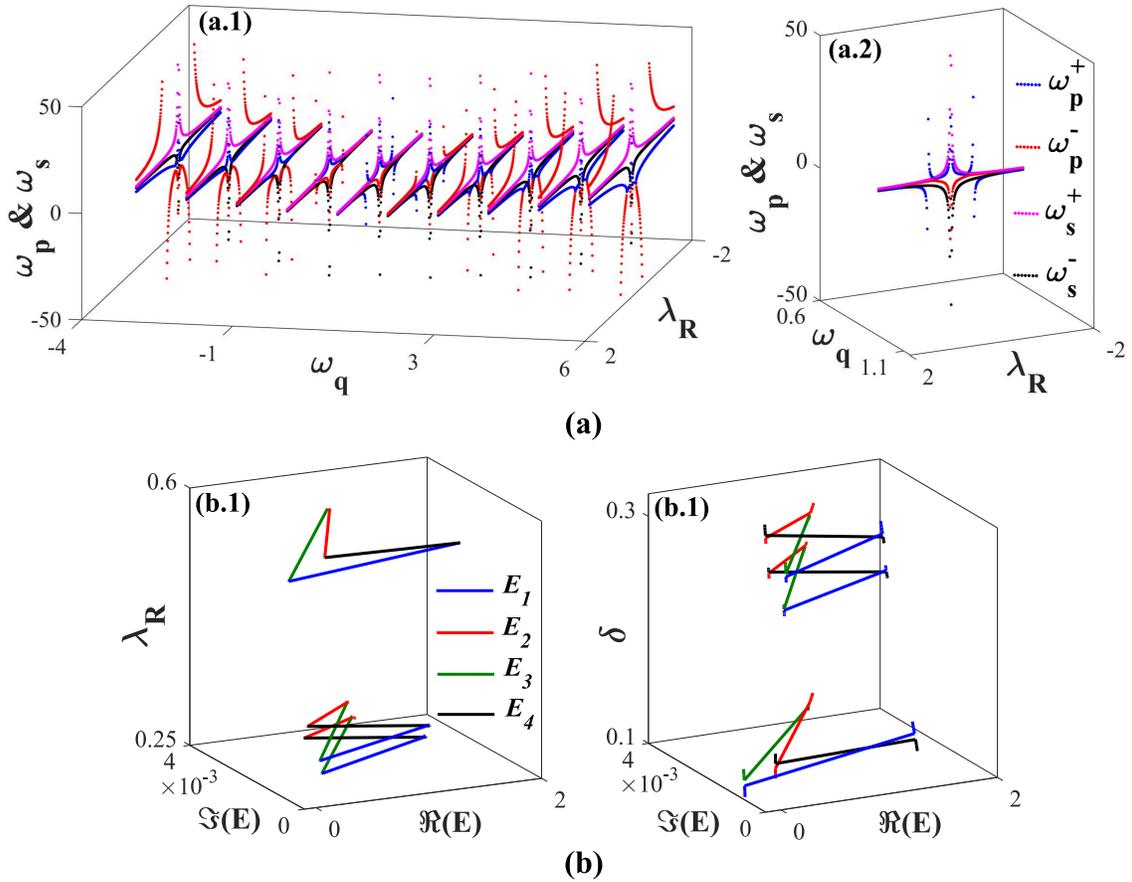}
	\caption{{\bf EP4-region and existence of multiple EP4}. \textbf{(a)} (a.1) Variation of perturbation parameters with respect to $\lambda_R$ forming a region that hosts multiple EP4. (a.2) A specific cross-section of the region shown in (a.1). \textbf{(b)} Existence of multiple EP4 governing the interactions between all the coupled states with respect to (b.1) $\lambda_R$ and (b.2) $\delta$ inside the region shown in (a.1).}
	\label{fig6}
\end{figure*}
Now, considering the four-fold coalescence at an EP4, we rigorously assume a {\it critical eigenvalue} which is the mean of all passive elements, at the coalescing point as
\begin{equation}
E_c=\frac{1}{4}\left(\widetilde{\varepsilon}_1+\widetilde{\varepsilon}_2+\widetilde{\varepsilon}_3+\widetilde{\varepsilon}_4\right)
\label{equation_Ec}
\end{equation}
which must satisfy the Eq. \ref{equation_Ea}. Again, extracting the $\omega_p$-terms from $p'_2,\,p'_3\,\textnormal{and}\,p'_4$ as
\begin{subequations}
	\begin{eqnarray}
	p'_2=p''_2-\lambda^2\omega_p^2,
	\label{equation_p11}\\
	p'_3=p''_3+\lambda^2\left(\widetilde{\varepsilon}_3+\widetilde{\varepsilon}_4\right)\omega_p^2,
	\label{equation_p22}\\
	p'_4=p''_4-\lambda^2\widetilde{\varepsilon}_3\widetilde{\varepsilon}_4\omega_p^2-\lambda^4(\omega_p^2+2\omega_p\omega_q\omega_r),
	\label{equation_p33}
	\end{eqnarray}
\end{subequations}

we can rewrite the Eq. \ref{equation_Ea} as
\begin{equation}
\mu_1\omega_p^2+\mu_2\omega_p+\mu_3=0;
\label{equation_omegap}
\end{equation}

with      
\begin{subequations}\begin{eqnarray}
\mu_1=\lambda^2\left(\widetilde{\varepsilon}_3+\widetilde{\varepsilon}_4-\widetilde{\varepsilon}_3\widetilde{\varepsilon}_4-\lambda^2-E_c^2\right),
\label{equation_mu1}\\
\mu_2=-2\lambda^4\omega_q\omega_r,
\label{equation_mu2}\\
\mu_3=E_c^4+p_1E_c^3+p''_2E_c^2+p''_3E_c+p''_4,
\label{equation_mu3}
\end{eqnarray}\end{subequations}

Here, $\left\{p''_2,p''_3,p''_4\right\}$ represent the parameters $\left\{p'_2,p'_3,p'_4\right\}$ after extraction of $\omega_p$-terms. Now, if we write $\omega_r$ in terms of $\omega_q$ using the relation $\omega_r=0.95-(\omega_q+0.1)/2$ (from Eq. \ref{equation_omega1} and \ref{equation_omega2} ) then Eq. \ref{equation_omegap} becomes a pure quadratic equation of $\omega_p$ having two different roots, say $\omega_p^+$ and $\omega_p^-$. 

Now, we consider a different special setting in Eq. \ref{equation_H} as $\omega_p=1$ to explore the relation of $\omega_s$ with $\omega_q$ and $\omega_r$ over the independent variation of $\lambda$. Here, we rewrite Eq. \ref{equation_H} as
\begin{equation}
\mathcal{H}\Bigg|_{\omega_p=1}=\left(\begin{array}{cccc}\widetilde{\varepsilon}_1 & 0 & 0 & \omega_q\lambda \\0 & \widetilde{\varepsilon}_2 & \omega_r\lambda & 0 \\0 & \omega_r\lambda & \widetilde{\varepsilon}_3 & \omega_s\lambda\\\omega_q\lambda & 0 & \omega_s\lambda & \widetilde{\varepsilon}_4\end {array}\right)+\lambda\left(\begin{array}{cccc}0 & 1 & 0 & 0 \\1 & 0 & 0 & 0 \\ 0 & 0  & 0 & 0\\ 0 & 0 & 0 & 0\end {array}\right).
\label{equation_H2}
\end{equation}
Here also considering the critical eigenvalue $E_c$ (given by Eq. \ref{equation_Ec}) and the relation between $\omega_r$ and $\omega_q$ as $\omega_r=0.95-(\omega_q+0.1)/2$ (from Eq. \ref{equation_omega2}), we can derive a pure quadratic equation of $\omega_s$ having the form 
\begin{equation}
\nu_1\omega_s^2+\nu_2\omega_s+\nu_3=0.
\label{equation_omegas}
\end{equation}
Here, the expressions of the terms $\nu_1,\,\nu_2\,\textnormal{and}\,\nu_3$ can be obtained in a similar way which is described for the previous special setting. Eq. \ref{equation_omegas} have two different roots, say, $\omega_s^+$ and $\omega_s^-$.

Now, we plot the roots $\left\{\omega_p^+,\omega_p^-\right\}$ (coming from Eq. \ref{equation_omegap}; represented by dotted blue and red curves, respectively) and $\left\{\omega_s^+,\omega_s^-\right\}$ (coming from Eq. \ref{equation_omegas}; represented by dotted magenta and black curves, respectively) in Fig. \ref{fig6}(a.1) for a continuous variation of $\lambda_R$ within [$-2, 2$] (with simultaneous variation of $\lambda_I$ maintaining the ratio $\lambda_I/\lambda_R=-10^{-3}$), taking different values of $\omega_q$. Here, we investigate the intersection region between the trajectories of these four roots. As can be seen in Fig. \ref{fig6}(a.1), we observe that for $\omega_q=-3.14$, there is no blank area under the intersections. Then, for an increasing $\omega_q$, the area under the intersections increases up to a specific value of $\omega_q=0.86$ and then again decreases even we increase $\omega_q$ further. For $\omega_q=5.86$, again there is no blank area under the intersections. Now, if we consider the overall range of $\omega_q$ from $-3.14$ to $5.86$ then we can realize a closed 3$D$-space in Fig. \ref{fig6}(a.1). A particular cross-section of this closed 3$D$-space, where area under the intersections becomes maximum (i.e., for $\omega_q=0.86$), has been shown in Fig. \ref{fig6}(a.2). 

We have named this closed 3$D$-space as ``EP4-region'' because within this region the proper coupling between the perturbation parameters $\omega_p,\,\omega_q,\,\omega_r\,\textnormal{and}\,\omega_s$ through $\delta$ for a continuous variation of $\lambda$ will happen to control the fourth-order interactions between four coupled states of the proposed Hamiltonian $\mathcal{H}$ (given by Eq. \ref{equation_H}). Thus in Fig. \ref{fig6}(a.1), we have shown the relation between $\omega_p$ and $\omega_s$ with $\omega_q$ and $\omega_r$ for a wide range of $\lambda$ where $\omega_r$ has been expressed in terms of $\omega_q$. If we express $\omega_q$ in terms of $\omega_r$ then also we can get a similar region as shown in Fig. \ref{fig6}(a.1); but in this case, $\omega_q$-axis will be replaced by $\omega_r$-axis. Note that, in this calculation, we consider two special settings by choosing a specific pair $\left\{\omega_p,\omega_s\right\}$ from four perturbation parameters. In a very similar way one can choose a different pair from the possible combinations to formulate such an EP4-region. Winding around this parametric region shown in Fig. \ref{fig6}(a.2), we encounter three different situations for the proposed Hamiltonian $\mathcal{H}$ where four coupled states are mutually interacting around three different EP4s. In Fig. \ref{fig6}(b.1), we have shown such three fourth-order interactions within the EP4-region with respect to $\lambda_R$. For proper validation, we have shown the same interactions between four coupled states with respect to $\delta$.       

\section{Conclusion}

In summary, for the first time, we have successfully reported the existence of a fourth-order exceptional point (EP4) by considering a four-level non-Hermitian Hamiltonian. Within the proposed framework, a passive system, hosting four decaying states, is subjected to a parameter-dependent perturbation. We have chosen a complex ($\lambda$) and a real ($\delta$) control parameters in such a way, that the system can host different orders of interaction phenomena between the coupled states in the vicinity of different orders of singularities. Here we have shown the simultaneous existence of EP2, EP3 and EP4 within a certain parametric range. Verifying the state-exchange phenomena between two and three coupled states around an EP2 and an EP3 respectively, we have exclusively established a successive state-conversion phenomenon between four coupled states following a parametric variation around an EP4. Introducing random fluctuation in the parametric variation around an EP4, the immutability of this successive state-conversion phenomena has been shown. We have also established that the topological properties of an EP of a specific order are robust even in the presence of another EPs of lower order inside the parametric loop. To co-relate the multiple locations of EP4s in a system, we have formulated an EP4-region by interplaying the specific relationship of perturbation parameters with the coupling control parameters. The chiral behavior of state-exchange phenomenon around EP4 has also been established. The systems realized with such exclusively proposed scheme may open up a fertile platform to improve the quality of a wide range of EP-aided state-of-the-art applications like all-optical mode conversions, optical sensing with enhanced sensitivity, etc. Owing to unconventional richer physical aspects, an EP4 should offer itself as a new light manipulation tool in integrated circuits.       

\section*{Acknowledgment} 

SB acknowledges the support from from Ministry of Human Resource Development (MHRD), India. AL and SG acknowledge the financial support from the Science and Engineering research Board (SERB), Ministry of Science and Technology, India under Early Career Research Scheme [Grant No. ECR/2017/000491].

\end{document}